\documentclass[10pt,twocolumn,letterpaper]{article}

\usepackage{cvpr}              

\usepackage[dvipsnames]{xcolor}

\definecolor{cvprblue}{rgb}{0.21,0.49,0.74}
\usepackage[pagebackref,breaklinks,colorlinks,citecolor=cvprblue]{hyperref}
\usepackage{caption}
\usepackage{subcaption}
\usepackage{amsmath}
\usepackage{amssymb}
\usepackage{nicefrac}

\def\paperID{16746} 
\def\confName{CVPR}
\def\confYear{2024}

\title{Segment Any Cell: A SAM-based Auto-prompting Fine-tuning Framework for Nuclei Segmentation}

\author{Saiyang Na \and Yuzhi Guo \and Feng Jiang \and Hehuan Ma \and Junzhou Huang$^{\ast}$\\
University of Texas at Arlington\\
Arlington, TX\\
Corresponding to: sxn3892@mavs.uta.edu $\&$ jzhuang@uta.edu\\
}

\begin{document}
\maketitle
\begin{abstract}
In the rapidly evolving field of AI research, foundational models like BERT and GPT have significantly advanced language and vision tasks. The advent of pretrain-prompting models such as ChatGPT and Segmentation Anything Model (SAM) has further revolutionized image segmentation. 
However, their applications in specialized areas, particularly in nuclei segmentation within medical imaging, reveal a key challenge: the generation of high-quality, informative prompts is as crucial as applying state-of-the-art (SOTA) fine-tuning techniques on foundation models. 
To address this, we introduce Segment Any Cell (SAC), an innovative framework that enhances SAM specifically for nuclei segmentation.
SAC integrates a Low-Rank Adaptation (LoRA) within the attention layer of Transformer to improve the fine-tuning process, outperforming existing SOTA methods. It also introduces an innovative auto-prompt generator that produces effective prompts to guide segmentation, a critical factor in handling the complexities of nuclei segmentation in biomedical imaging.
Our extensive experiments demonstrate the superiority of SAC in nuclei segmentation tasks, proving its effectiveness as a tool for pathologists and researchers. Our contributions include a novel prompt generation strategy, automated adaptability for diverse segmentation tasks, the innovative application of Low-Rank Attention Adaptation in SAM, and a versatile framework for semantic segmentation challenges.

\end{abstract}
\section{Introduction}
\label{sec:intro}

In recent years, the landscape of AI research has undergone a profound transformation, primarily driven by the utilization of extensive datasets for large-scale model training. This paradigm shift has given rise to foundation models, exemplified by renowned examples like BERT~\cite{devlin2018bert}, GPT~\cite{brown2020language}, and ViT~\cite{dosovitskiy2020image} which have illustrated remarkable proficiency in a diverse range of language and vision-related tasks~\cite{bommasani2021opportunities, yu2022coca}. 
The breakthrough launch of ChatGPT introduces the pretrain-prompting approach. It innovatively leverages an exceptionally large language foundation model and allows users to customize its application through personalized prompts tailored to their specific tasks.
Inspired by such, Segmentation Anything Model (SAM)~\cite{kirillov2023segment} has gained substantial attention for its exceptional capabilities as a versatile vision segmentation model.
SAM distinguishes itself by its capacity to generate a wide array of finely detailed segmentation masks, all guided by user prompts. This emergence of SAM signifies a revolutionary advancement in the realm of image segmentation and related fields within computer vision, promising innovative and robust solutions for various applications~\cite{tang2023can,huang2023segment,ji2023sam}.

While SAM has shown promise in various domains, its applicability is not all-encompassing for medical image-related tasks. This limitation is consistent with other foundation models, as the training data cannot comprehensively represent the entirety of potential scenarios within computer vision~\cite{bommasani2021opportunities,nori2023capabilities,lee2023benefits,raffel2020exploring}. Fortunately, recent advancements in fine-tuning and Low-Rank Adaptation (LoRA) methods for large language models~\cite{hu2021lora, zhang2023llama, touvron2023llama} have enabled some improvements in SAM's performance on downstream medical image tasks~\cite{wu2023medical, chen2023sam, ma2023segment}. For example, ``Medical SAM Adapter'' (MSA) presents an adapter to integrate domain knowledge on multi-organ and polyp segmentation~\cite{wu2023medical}.
However, the unique challenges presented by nuclei segmentation, a crucial subfield of biomedical image analysis, require special attention. Nuclei serve as the fundamental building blocks of life, and play a pivotal role in digital histopathology image analysis~\cite{sharma2022mani, yang2021minimizing}. Yet, accurate nuclei segmentation presents a demanding challenge due to the diverse and complex nature of nuclei. These complexities include variations in shape, appearance, clustering, overlap, blurred boundaries, inconsistent staining methods, scanning artifacts, etc. Furthermore, different organs and cancer types in histopathology may exhibit distinct textures, color distributions, morphology, and scales, adding an additional layer of complexity to the segmentation tasks~\cite{xu2017large, mahmood2019deep}.



\begin{figure*}[htbp]
     \centering
     \begin{subfigure}[t]{0.286\linewidth}
         \centering
         \includegraphics[width=\linewidth]{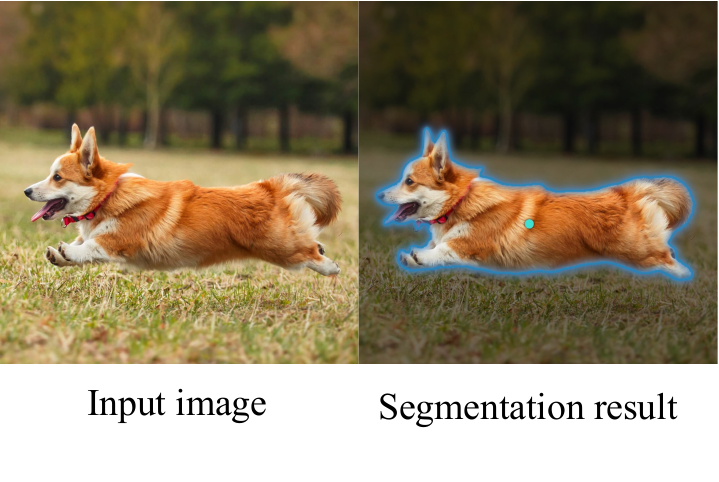}
         \caption{Natural Image Segmentation.}
         \label{fig:intro_demo_a}
     \end{subfigure}
     \begin{subfigure}[t]{0.702\linewidth}
         \centering
         \includegraphics[width=\linewidth]{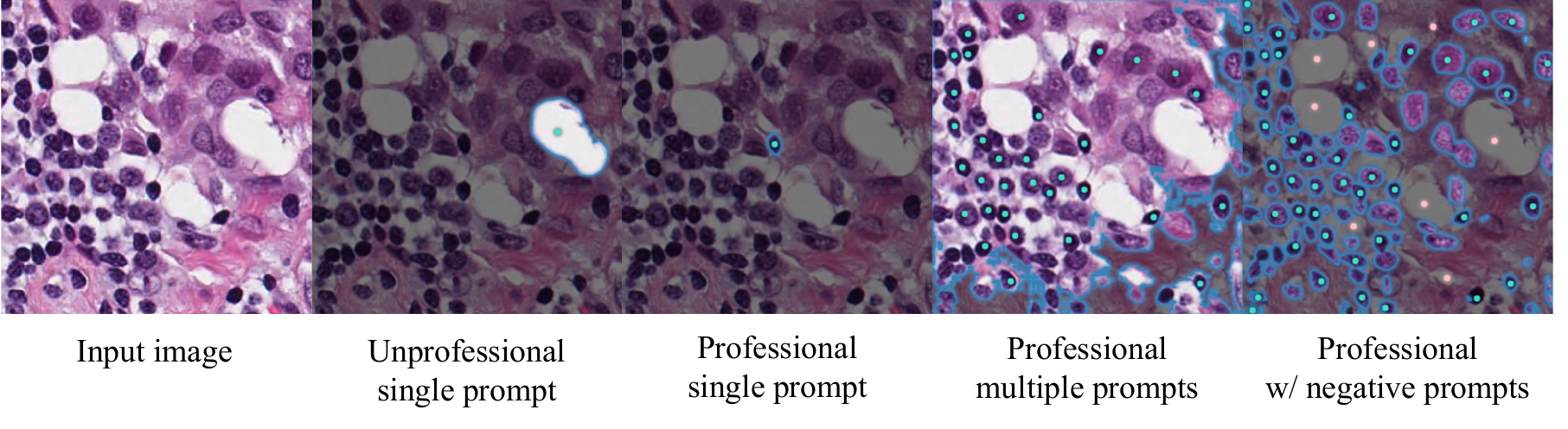}
         \caption{Nuclei Segmentation.}
         \label{fig:intro_demo_b}
     \end{subfigure}
        \caption{Comparative segmentation results using the SAM demo: (a) Segmentation results for natural images using SAM. (b) Segmentation results for cell nucleus images using SAM with different prompting strategies: Unprofessional single prompt – A single random prompt provided by a non-expert, potentially acting as a noisy prompt leading to failed segmentation results; Professional single prompt – A single positive prompt given by a professional expert, such as a pathologist; Professional multiple prompts – Multiple positive prompts provided by a professional; Professional w/ negative prompts – A few negative prompts provided by a professional. The blue dots represent the given positive prompts, while the pink dots represent the negative prompts.}
        \label{fig:intro_demo}
\end{figure*}

Therefore, we argue that merely fine-tuning SAM is insufficient for cell nuclei segmentation. 
We propose exploring the quality of prompts to enhance segmentation performance. Figure \ref{fig:intro_demo} compares the segmentation results obtained by using SAM directly with those achieved through professionally annotated prompts. As observed in Figure \ref{fig:intro_demo_a}, SAM correctly segments the puppy when the center of the object is clicked. However, when an unprofessional single prompt is used by clicking on the input nuclei image, as depicted in Figure \ref{fig:intro_demo_b}, the result is incorrect due to the lack of sufficient domain knowledge. 
Conversely, when a professional single prompt is precisely placed on a cell nucleus, SAM is only able to correctly segment that particular cell. Nevertheless, it is important to note that an increase in the number of prompts can sometimes lead to decreased model performance. As illustrated in Figure \ref{fig:intro_demo_b} - Professional multiple prompts, most areas are segmented as a whole when multiple prompts are provided. Therefore, negative prompts are crucial for guiding accurate segmentation. The ``Professional w/ negative prompts'' setting effectively distinguishes between most nuclei and non-nuclei.
Thus, consistent with our assumption, a well-crafted prompt significantly enhances SAM's capabilities.
In other words, providing SAM with sufficiently informative prompts that help the model distinguish between cell and non-cell regions can enhance its performance. Additionally, this implementation can be seen as a beneficial add-on, easily integrating with existing fine-tuning or adaptation state-of-the-art (SOTA) methods without any conflicts.
The next challenge is how to provide or obtain qualified prompts. Given that a digital microscopic tissue image typically contains hundreds of nuclei, manually pointing out all the correct nuclei is impractical. Therefore, developing methods to automatically generate such nuclei prompts, or to provide a limited number of examples, is becoming essential.
In this paper, we present Segment Any Cell (SAC), an automatic prompting-based fine-tuning framework for nuclei segmentation, which is built upon the SAM model. 
Specifically, we propose implementing a LoRA on the attention layer within the transformer to enhance the fine-tuning process of SAM, which outperforms existing fine-tuning/adapter-based SOTA methods. We also designed an auto-prompt generator that can automatically generate a large number of high-quality prompts to guide the segmentation process. This addresses the limitation of SAM's difficulty in obtaining suitable prompts for nuclei segmentation tasks.  
Through extensive experiments, we demonstrate the superiority of our method. The contributions of this paper are summarized as follows:
\begin{enumerate}
  \item We propose a novel nuclei prompt generation and discriminating strategy on SAM, significantly enhancing nuclei segmentation performance. This approach is complementary to existing SOTA methods that fine-tune the encoder.
  \item Our proposed SAC model is fully automated and easily adaptable to various nuclei segmentation tasks, offering a simplified yet effective tool for pathologists and researchers.
  \item We introduce an innovative application of a Low-Rank Attention Adaption in SAM, substantially advancing nuclei segmentation. This method surpasses existing fine-tuning/adaptation techniques for large models without increasing the number of parameters, thereby improving both precision and efficiency in segmentation tasks.
  \item Our approach presents a general fine-tuning framework that is versatile for different semantic segmentation tasks based on SAM, 
  which share similarities with the challenges faced in nuclei segmentation.
\end{enumerate}

\section{Related Works}
\label{sec:related}

\subsection{Nuclei segmentation}
In the domain of nuclei segmentation, supervised methods have been widely employed, encompassing classical techniques such as U-Net~\cite{ronneberger2015u}, Mask R-CNN~\cite{he2017mask}, and FCN~\cite{long2015fully}. However, these methods often necessitate substantial labeled training data. 
In contrast, recent advancements have explored the use of pre-trained models trained on medical images~\cite{haq2022self, lee2022cellseg}. Nevertheless, a critical challenge emerges from the scarcity of nuclei image data for pre-training, limiting the potential of models exclusively pre-trained on nuclei data. Thus, integrating models pre-trained on large-scale natural image datasets~\cite{he2016deep, simonyan2014very, dosovitskiy2020image, he2022masked, kirillov2023segment}, such as ImageNet~\cite{deng2009imagenet}, into the nuclei segmentation pipeline presents a crucial open challenge in the era of large models, requiring innovative strategies for techniques of efficient fine-tuning of large models. 

\subsection{Segment Anything Model}
Segment Anything Model (SAM)~\cite{kirillov2023segment} is trained on a large visual corpus and demonstrates remarkable segmentation capabilities across diverse scenarios. Its key innovation lies in its ability to segment a wide range of objects or regions in images based on user-defined prompts, making it a pioneering approach towards addressing the challenges of generic image segmentation. 
However, it is worth noting that SAM, despite its versatility and success in generic image segmentation, has certain limitations when applied to the medical imaging domain. One notable drawback is the insufficient quantity of training data explicitly collected for medical usage. While SAM's training process incorporates a sophisticated and efficient data engine, it has gathered a relatively small number of medical image cases. This limitation poses challenges when adapting SAM for tasks in medical image analysis, where specialized datasets and domain knowledge are crucial for achieving optimal segmentation results.
The work proposed in this paper aims to leverage SAM to solve downstream nuclei segmentation tasks.

\subsection{Fine-tuning and Adaptation on Large Models}

The fine-tuning technologies for large models like LoRA~\cite{hu2021lora} and Llama~\cite{touvron2023llama} are designed for efficient and specialized adaptation of language models. LoRA adjusts a minimal number of parameters, enabling specific customizations without heavy re-training. Llama offers both a versatile foundation model and fine-tuned versions for various tasks.
Applying similar technology to SAM-based medical image fine-tuning involves tailoring SAM for more specialized applications. MSA~\cite{wu2023medical} focuses on adapting SAM for precise medical image segmentation, enhancing its ability to interpret complex medical imagery. Meanwhile, ``SAM-Adapter''~\cite{chen2023sam} targets SAM's limitations in challenging scenarios like camouflage or shadowy environments, thereby extending its applicability and performance in varied and difficult conditions. These adaptations signify a trend towards more efficient, targeted, and effective use of SAM in specialized image analysis tasks. 

However, these fine-tuning methods often overlook a crucial aspect of the SAM: the choice of prompts. The effectiveness of SAM heavily relies on the quality and specificity of the prompts provided, especially in tasks like cell nucleus segmentation. A poorly chosen prompt can lead to suboptimal results, as the model might not fully grasp the intricacies of distinct tasks. This emphasizes the need for more refined and contextually relevant prompt engineering to fully harness SAM's capabilities in specialized tasks such as medical image analysis. The method developed in this study aims to provide SAM with better prompts while also achieving automated prompting. This facilitates a more convenient and efficient way to enhance its performance in precise and complex segmentation tasks, such as cell recognition, thereby addressing a key limitation in current fine-tuning methods.

\section{Method}
\label{sec:method}

\begin{figure*}[htbp]
    \begin{center}
        \includegraphics[width=0.75\linewidth]{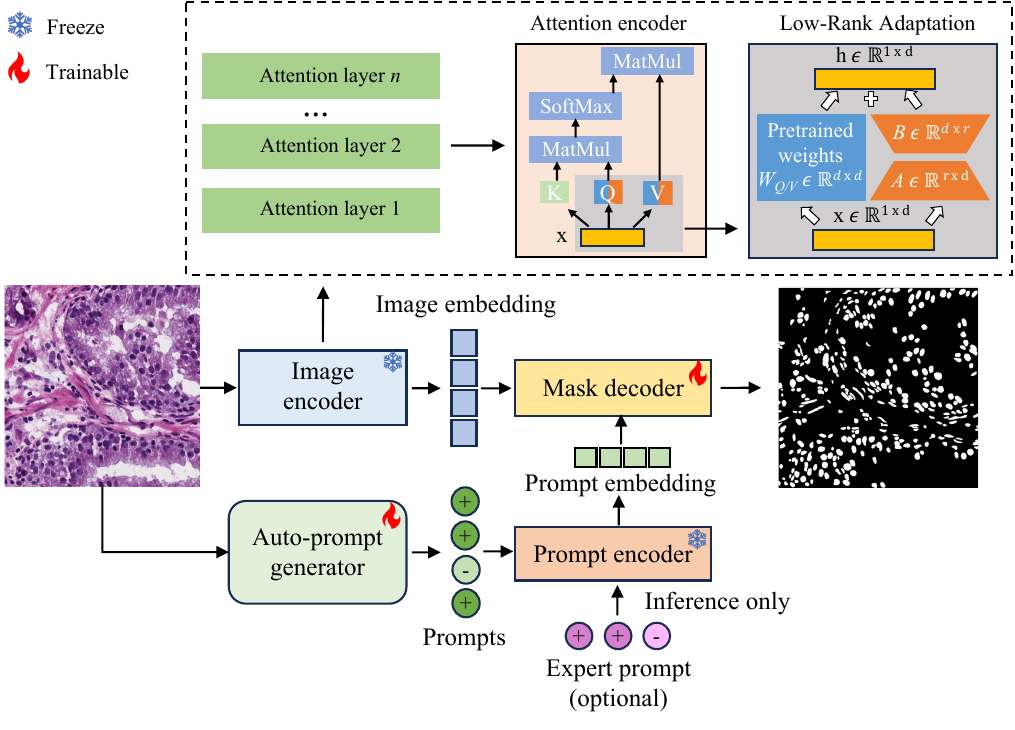}
    \end{center}
    \caption{Overall framework of SAC. The medical images are initially fed into the frozen parametered SAM image encoder, where we apply a Low-Rank attention adapter on each attention layer for efficient generation of image embeddings. Concurrently, images are processed through our innovative auto-prompt generator, producing both positive and negative prompts. These prompts are then fed into SAM prompt encoder to obtain prompt embeddings. Last, both image and prompt embeddings are input into a trainable SAM mask decoder (fine-tuning) to produce the final segmentation results. Notably, during the inference phase, our framework also allows for manual input of prompts as an option, potentially aiding in the segmentation of cell nuclei.}
    \label{fig:framework}
\end{figure*}

\subsection{Overview}

The overall framework of our proposed SAC is illustrated in Figure \ref{fig:framework}. We introduce an innovative model that enhances the Segment Anything Model (SAM) core framework for cell nuclei segmentation tasks. Specifically, we have augmented the SAM architecture by incorporating a Low-Rank Adaptation (LoRA) approach into a Vision Transformer (ViT). This integration enables a more effective and efficient fine-tuning process for SAM, substantially improving the model's adaptability and performance in cell nuclei segmentation tasks. Furthermore, we introduce an ``Automatic Prompt Generator'' designed to automatically produce a high quality of contextually relevant prompts. This addresses the inherent limitations of SAM in cell nuclei segmentation, thereby increasing segmentation accuracy and efficiency.

\subsection{Low-Rank Adaptation for SAM Image Encoder}

Our model introduces a novel implementation of the LoRA within the SAM, focusing specifically on the internal modifications of the transformer architecture. 
This unique approach significantly differentiates our model from traditional methods of incorporating adapters in transformers.
In contrast to typical adapters that add trainable parameters between transformer layers, our implementation of LoRA directly modifies the internal structure of the transformer, which is the Query-Key-Value (QKV)  \cite{vaswani2017attention} matrices in the attention layers. 
The innovation in our approach lies in the selective adaptation of the Q and V matrices, while keeping the K matrix unchanged. 
This focused modification allows for a more controlled and effective adaptation process, which can be formulated as: 
\begin{equation}
    h = W_{0}x + W_{\Delta}x = W_{Q/V}x + BAx,
\end{equation}
where \(h\) represents the output of the adapted layer,  \(W_{0}\) denotes the original weight matrix associated with either the Q or V matrix in the attention layer, and $x \in \mathbb{R}^{1 \times d}$ is the input to the layer. \(W_{\Delta} = BA\) represents the Low-Rank adaptation applied to the layer, where $B\in \mathbb{R}^{d\times r}$ and $A \in \mathbb{R}^{r\times d}$ are the Low-Rank matrices.
By altering the Q and V components, the LoRA-equipped SAM model achieves enhanced sensitivity and specificity in segmenting complex patterns.

\subsection{Auto-prompt Generator}\label{sec:apg}

In our enhanced SAM framework, while the LoRA plays a critical role in fine-tuning, its efficacy can be maximized when combined with precise prompts, particularly for complex tasks like segmenting cell nuclei. 
For the segmentation of nuclei, a single image may contain hundreds to thousands of cells, making manual prompting impractical. 
This is where our Auto Prompt Generator becomes indispensable.
Therefore, a prompt that can precisely guide segmentation without requiring significant labor efforts is urgently needed. 
Recognizing the limitations of manual prompt generation in such scenarios, our model incorporates an innovative component: auto-prompt generator.

Our proposed auto-prompt generator is an auxiliary neural network designed to automate the process of generating prompts. 
Let \(I\) be the input image, image \(I\) is passed through an auxiliary neural network \(\mathcal{F}\) (UNet is used in this study), and outputs a mask \(M\) as:
\begin{equation}
    M = \mathcal{F}(\theta_{u}, I),
\end{equation}
where \(\theta_{u}\) refer to the auxiliary neural network parameters.
The Binary Cross-Entropy (BCE) Loss is used to optimize the binary classification task performed by $\mathcal{F}$, which is:
\begin{equation}
\begin{aligned}
    \text{BCELoss}(M^{*}, M') =&  -\frac{1}{N} \sum_{i=1}^{N} [\ M'_{i} \log(M_{i}^{*}) \\ &
    + (1 - M'_{i}) \log(1 - M_{i}^{*})],
\end{aligned}
\end{equation}
where \( M^{*} = \text{sigmoid}(M) \) is the predicted probability (output of $\mathcal{F}$), \( M' \) is the true label (target), and \( N \) is the number of observations.

The integration of our auto-prompt generator in the SAM framework is further refined by the inclusion of prompt discrimination capabilities. This feature is crucial for cell nuclei segmentation as SAM's ability to distinguish between positive and negative prompts significantly impacts segmentation accuracy. As shown in Figure \ref{fig:intro_demo_b}, using only positive prompts often results in the unintentional selection of surrounding content, leading to inaccurate segmentation.

Our auto-prompt generator addresses this challenge by not only automating the generation of prompts but also discerning between positive and negative prompts. 
A binary classification task is then performed on the mask values \(M\) to determine the probability of each position being a positive or negative prompt as:
\begin{equation}
    P = (P_{pos}, P_{neg}) = \text{sigmoid}(M),
\end{equation}
where \(P_{pos}\) and \(P_{neg}\) represent the probabilities of positive and negative prompts, respectively.
From these probabilities, a certain number of points are selected based on a predefined probability threshold to serve as inputs for the prompt encoder:
\begin{equation}
    \mathcal{P} = \mathcal{S}(P),
\end{equation}
where \(\mathcal{S}\) represents the select prompts methods, and \(\mathcal{P}\) denotes the selected prompts.
These selected prompts are then used as the inputs to the prompt encoder.
For selecting the prompts from the probability, we introduce two methods: a centroid-based selection to generate \(\mathcal{P}_{O}\) and a direct probability-based selection to generate \(\mathcal{P}_{R}\).
In the centroid-based selection, connected regions of positive or negative probabilities are identified. 
The centroid of each region is then calculated and used as the prompt \(\mathcal{P}_{O} = \mathcal{O}(\mathcal{C}(P))\),
where \(\mathcal{O}\) is the function to calculate centroid, and \(\mathcal{C}\) denotes the function for calculating connection region.
This method ensures spatial coherence by focusing on the centroids of connected regions, leading to more contextually relevant prompts.

The alternative selection method involves directly randomly selecting points from the positive and negative probability with a threshold \(\tau\):
\(\mathcal{P}_{R} = \mathcal{R}(\tau, P)\),
where \(\mathcal{R}\) refers to the randomly selecting.
This method allows for more flexibility as it can select prompts from any part of the probability maps, not restricted to connected regions. Direct selection is also straightforward and can be faster as it does not require the computation of centroids or identification of connected regions.  
There is a possibility that the selected prompts may fall into boundary areas, which might affect accuracy. We have conducted ablation studies on the selection methods to compare the performance, and the details are provided in the section \ref{sec:cvd}.




Our proposed automated and discriminated prompting manner revolutionizes the segmentation process, especially in scenarios involving large numbers of cell nuclei. It ensures scalability, efficiency, and high accuracy in segmentation tasks, overcoming the limitations of manual prompt generation and the challenges posed by undiscriminated positive prompts.


\subsection{Prompt Encoder}
In our model, the prompt encoder from SAM is employed in a frozen state, so that we can focus exclusively on points prompts while disregarding masks and boxes. The points prompts can be derived from two sources:

\begin{itemize}
    \item automatically generated prompts, as detailed in auto-prompt generator in Section \ref{sec:apg}. We denote them as ``points'';
    \item professionally annotated points, which are manually selected for precision by experts, serve as the true labels. We denote them as ``experts''.
\end{itemize}
Such configurations allow the prompt encoder to process specific, targeted inputs, ensuring accuracy and relevance in the segmentation task.

\subsection{Mask Decoder}
We use the mask decoder from SAM to generate segmentation masks.
The mask decoder is not frozen since it is directly involved in the downstream segmentation task and constitutes a small portion of the overall parameters, so we keep it training.
It goes through normal fine-tuning as part of the neural network training process, which takes the prompt embeddings and image embeddings as inputs, and producing segmentation masks and scores as outputs.
\section{Experiments}
\label{sec:exp}

\subsection{Experiment set up}
\subsubsection{Datasets}
We conduct a series of extensive experiments on two datasets to evaluate the performance of our SAC: MoNuSeg~\cite{kumar2017dataset}, 
and the 2018 Data Science Bowl (DSB)~\cite{caicedo2019nucleus}. 
MoNuSeg is a nuclear segmentation dataset for digital microscopic tissue images. It includes 30 images for training and 14 images for testing. 
DSB is sourced from the 2018 Data Science Bowl challenge and serves to identify nuclei in diverse images. It contains 670 images, randomly split into 80\% for training, 10\% for validation, and 10\% for testing.
We follow the exact same settings for these datasets as those in the baseline papers for a fair comparison.  
For more extended datasets, please see the supplementary~\ref{suppl_Experiment_Detail}.

\subsubsection{Baselines}




To demonstrate the effectiveness of our model, we compare SAC with twelve baseline methods on the nuclei segmentation task. Specifically, eight of them are supervised learning methods: U-Net~\cite{ronneberger2015u}, UCTransNet~\cite{wang2022uctransnet}, MedT~\cite{valanarasu2021medical}, HistoSeg~\cite{wazir2022histoseg}, DuAT~\cite{tang2022duat}, SSFormer~\cite{wang2022stepwise}, MSRF-Net~\cite{srivastava2021msrf}, FANet~\cite{tomar2022fanet}; and three methods are based on pre-trained models: MDM~\cite{pan2023masked}, DoubleU-Net~\cite{jha2020doubleu}, and MSA~\cite{wu2023medical}. Different settings of SAM~\cite{kirillov2023segment} have also been experimented with for comparison.  
The detailed descriptions of the comparison methods are deferred in the Supplementary~\ref{suppl_baseline}.

\subsubsection{Training Details}

\textbf{Framework and Hardware:} 
Our experiments are conducted using \textit{PyTorch 2.1.1} on an \textit{Intel Xeon Gold 5420+} CPU and \textit{NVIDIA H100 PCIe} GPU with \textit{CUDA 12.3} and Driver Version \textit{545.23.06}. 

\textbf{Model Configuration:} 
The utilized base model is \textit{SAM vit-h}, and a 4-layer UNet is used as auto-prompt generator. 

\textbf{Loss Function:}~For the auto-prompt generator, \textit{BCELoss}~\cite{good1952rational} is employed as it is a binary classification task. The final segmentation results are evaluated using both \textit{Focal} \cite{lin2017focal} and \textit{DiceLoss} \cite{milletari2016v}, as per SAM's methodology. We also compare with the conventional \textit{DiceCELoss} in the supplementary~\ref{sec:lossc}.

\textbf{Performance Metrics and Training Procedure:} Models are evaluated based on \textit{F1 score}, \textit{Dice coefficient}, and \textit{Intersection over Union (IoU)}. 
We run a minimum of 30 epochs for fine-tuning with an early stop patience of 10, and then evaluate the segmentation performance on an independent test set.

\subsection{Experimental Results}
We compare the performance of our proposed SAC model on two widely recognized nuclei segmentation datasets, MoNuSeg and DSB, with commonly known SOTA nuclei segmentation methods, advanced large model with fine-tuning and adaptation techniques. 
The results are shown in Tables~\ref{tab:MoNuSeg} and~\ref{tab:DSB}. 
The first part of the baselines are either based on convolutional neural network (CNN) or transformer architectures, i.e., U-Net, UCTransNet, MedT, HistoSeg, DuAT, SSFormer, MSRF-Net, FANet, MDM, and DoubleU-Net. Next, we evaluate the performance with SAM 1-expert, which means only one manual prompt is given by an expert. SAM-FT represents the fine-tuned SAM with our datasets.
We also compare our model with a recent method, MSA, which is implemented based on a fine-tuned SAM with an adaptation strategy. Since SAM and MSA are not explicitly designed for nuclei segmentation, they do not contain results for our datasets. Therefore, we reproduce these methods and run the experiments on MoNuSeg and DSB to report the results. For other methods, we directly use the originally reported results in the papers since we follow their settings. The evaluation metrics include F1, IoU, and Dice score.

\begin{table}[!h]
  \centering
  \resizebox{0.8\linewidth}{!}{
  \begin{tabular}{@{}lccc@{}}
    \toprule
    \textbf{Method} & \textbf{F1} & \textbf{IoU} & \textbf{Dice} \\
    \midrule
    U-Net~\cite{ronneberger2015u} & 79.43 & 65.99 & - \\
    MedT~\cite{valanarasu2021medical} & 79.55 & 66.17 & - \\
    UCTransNet~\cite{wang2022uctransnet} & - & 65.5 & 79.08 \\
    MDM~\cite{pan2023masked} & - & - & 81.01 \\
    HistoSeg~\cite{wazir2022histoseg} & 75.08 & 71.06 & - \\
    DoubleUnet~\cite{jha2020doubleu} & - & 62.82 & 77.16 \\
    \midrule
    SAM 1-expert~\cite{kirillov2023segment}  & 25.36 & 14.24 & 24.03 \\
    SAM-FT 1-expert~\cite{kirillov2023segment}  & 81.57 & 68.76 & 81.40 \\
    \midrule
    MSA 1-expert~\cite{wu2023medical} & 81.65 & 69.07 & 81.62 \\
    \midrule
    \textbf{SAC 0-expert} & \textbf{84.11} & \textbf{72.61} & \textbf{84.03} \\
    \bottomrule
  \end{tabular}
  }
  \caption{Comparison results on MoNuSeg dataset using F1, IoU and Dice scores, where higher values indicate better performance. Best results are highlighted as \textbf{bold}. }
  \label{tab:MoNuSeg}
\end{table}



\begin{table}[!h]
  \centering
  \resizebox{0.8\linewidth}{!}{
  \begin{tabular}{@{}lccc@{}}
    \toprule
    \textbf{Method} & \textbf{F1} & \textbf{IoU} & \textbf{Dice} \\
    \midrule
    U-Net~\cite{ronneberger2015u} &-  & 83.10 & 90.80 \\
    UCTransNet~\cite{wang2022uctransnet} &  -& 83.50 & 91.10 \\
    DoubleUnet~\cite{jha2020doubleu} &  -& 84.07 & 91.33 \\
    DuAT~\cite{tang2022duat} & - & 87.00 & 92.60 \\
    SSFormer-L~\cite{wang2022stepwise} & - & 86.14 & 92.30 \\
    MSRF-Net~\cite{srivastava2021msrf} & - & 85.34 & 92.24 \\
    FANet~\cite{tomar2022fanet} & 91.76 & 85.69 & - \\
    \midrule
    SAM 1-expert~\cite{kirillov2023segment} & 66.79 & 58.37 & 69.79 \\
    SAM-FT 1-expert~\cite{kirillov2023segment} & 92.89 & 86.17 & 92.37 \\
    \midrule
    MSA  1-expert~\cite{wu2023medical} & 93.24 & 86.94 & 92.85 \\
    \midrule
    \textbf{SAC 0-expert} & \textbf{93.48} & \textbf{87.32} & \textbf{93.04} \\
    \bottomrule
  \end{tabular}
  }
  \caption{Comparison results on DSB dataset using F1, IoU and Dice scores, where higher values indicate better performance. Best results are highlighted as \textbf{bold}. }
  \label{tab:DSB}
\end{table}




As shown in Table~\ref{tab:MoNuSeg} and~\ref{tab:DSB}, the zero-shot performance of SAM (SAM 1-expert) is generally worse than that of fully-trained models on the target nuclei segmentation tasks, regardless of the prompt given. To ensure a fair comparison, we also include results from a fine-tuned version of SAM (SAM-FT), specifically trained on the dataset utilized in this study. This observation reveals the limited zero-shot transfer capability of SAM in medical imaging contexts, and highlights that even with fine-tuning, SAM struggles to achieve SOTA performance. This phenomenon aligns with findings from several other research studies~\cite{wu2023medical, deng2023segment, ma2023segment, he2023accuracy, roy2023sam}. 
Moreover, while MSA applies advanced SAM adaptation techniques on top of SAM-FT, it achieves only marginal improvements. In contrast, our method demonstrates further enhancements, which highlight the superiority of our approach by integrating an auto-prompt generator and low-rank adaptation within our nuclei prediction framework. These findings emphasize the effectiveness of our novel contributions in enhancing the performance of SAM-based segmentation tasks.

Overall, our method achieves superior performance compared to SOTA cell nuclei segmentation methods, outperforming the current advanced fine-tuning SAM (SAM-FT) and SAM adaptation methods (MSA). Importantly, our method enables the automatic generation of prompts, eliminating the need for expert input, and demonstrating a blend of superior performance and ease of use.

\subsection{Ablation Studies}

\subsubsection{Effectiveness of Prompts in SAM Fine-Tuning}

We conduct ablation studies to assess the impact of prompt quality. We employ different models as the backbone, comparing the results with varying numbers of auto-prompts (point) and manual-prompts (expert). For example, ``SAM 0-point 3-expert'' represents SAM model with zero auto-prompt and three manual prompts. 

\begin{table}[htbp]
\centering
\resizebox{0.85\linewidth}{!}{
\begin{tabular}{@{}lccc@{}}
\toprule
\textbf{Method} & \textbf{F1} & \textbf{IoU} & \textbf{Dice} \\
\midrule
SAM 0-point 0-expert & 26.29 & 13.26 & 22.73 \\
SAM 0-point 3-expert & 27.05 & 15.35 & 26.32 \\
\midrule
SAM-FT 0-point 0-expert & 79.78 & 68 & 80.89 \\
SAM-FT 1-point 0-expert & 80.81 & 68.31 & 81.07 \\
SAM-FT 1-point 1-expert & 81.57 & 68.76 & 81.4 \\
SAM-FT 1-point 3-expert & 81.39 & 68.5 & 81.24 \\
SAM-FT 3-point 3-expert & 81.73 & 69.04 & 81.6 \\
\midrule
MSA 0-point 0-expert & 81.57 & 69.39 & 81.61 \\
MSA 0-point 1-expert & 81.65 & 69.07 & 81.62 \\
MSA 1-point 0-expert & 81.65 & 69.12 & 81.66 \\
MSA 1-point 1-expert & 83.78 & 72.14 & 83.73 \\
MSA 3-point 3-expert & 84.03 & 72.53 & 84.01 \\
\midrule
LoRA 0-point 0-expert & 84.03 & 72.52 & 84.01 \\
LoRA 0-point 1-expert & 83.95 & 72.46 & 83.98 \\
LoRA 1-point 0-expert & 84.03 & 72.61 & 84.11 \\
LoRA 1-point 1-expert & 84.09 & 72.67 & 84.18 \\
\textbf{LoRA 3-point 3-expert} & \textbf{84.35} & \textbf{72.94} & \textbf{84.29} \\
\bottomrule
\end{tabular}
}
 \caption{Performance of different prompt settings on MoNuSeg Dataset using various backbone models. ``SAM'' and ``MSA'' represent SAM and MSA models. ``-FT'' means SAM with fine-tuning. ``LoRA'' is SAM with our LoRA implementation. The ``-point'' and ``-expert'' annotations represent the number of auto-generated and expert prompts used, respectively.}
\label{tab:full_prompt_effectiveness}
\end{table}

Table~\ref{tab:full_prompt_effectiveness} shows that
models with few-expert learning (1-point or 3-point with experts) generally outperform the models without fine-tuning and those with zero-expert learning. This suggests that including a small number of expert prompts can guide the SAM model toward better segmentation performance.
The models with LoRA as the backbone tend to achieve higher Dice and IoU scores compared to MSA, indicating the efficacy of LoRA in this context.
Moreover, incremental increases in the number of auto-generated prompts (points) with expert prompts show slight improvements, demonstrating that combining automated and expert guidance benefits performance. The comparison between ``SAM'' and ``SAM-FT'' illustrates that fine-tuning with expert input may provide the model with more accurate and relevant features to learn from, leading to better generalization on the segmentation task. 
Furthermore, the superior performance of LoRA models might be attributed to their ability to better preserve and utilize information from the pre-trained model during fine-tuning. 

\begin{table}[ht]
\centering
\resizebox{1.0\linewidth}{!}{
\begin{tabular}{@{}c|ccc|ccc@{}}
\toprule
\textbf{Method} & \multicolumn{3}{c|}{\textbf{MoNuSeg}} & \multicolumn{3}{c}{\textbf{DSB}} \\
\cmidrule{2-7}
 & \textbf{F1} & \textbf{IoU} & \textbf{Dice} & \textbf{F1} & \textbf{IoU} & \textbf{Dice} \\
\midrule
SAM-FT 0-point 0-expert & \textbf{79.78} & \textbf{68} & \textbf{80.89} & \textbf{92.89} & \textbf{86.17} & \textbf{92.37} \\
SAM-FT 0-point 1-expert & 63.91 & 46.67 & 63.13 & 91.41 & 84 & 90.97 \\
SAM-FT 0-point 3-expert & 59.5 & 41.83 & 57.47 & 90.77 & 82.52 & 90.07 \\
\bottomrule
\end{tabular}
}
\caption{\label{tab:decoder-results} Comparison results of fine-tuned SAM w/o auto-prompts on MoNuSeg and DSB datasets, using different numbers of manual expert prompts, i.e., 0, 1, 3. }
\end{table}

\begin{figure}[!h]
    \centering
    \includegraphics[width=1\linewidth]{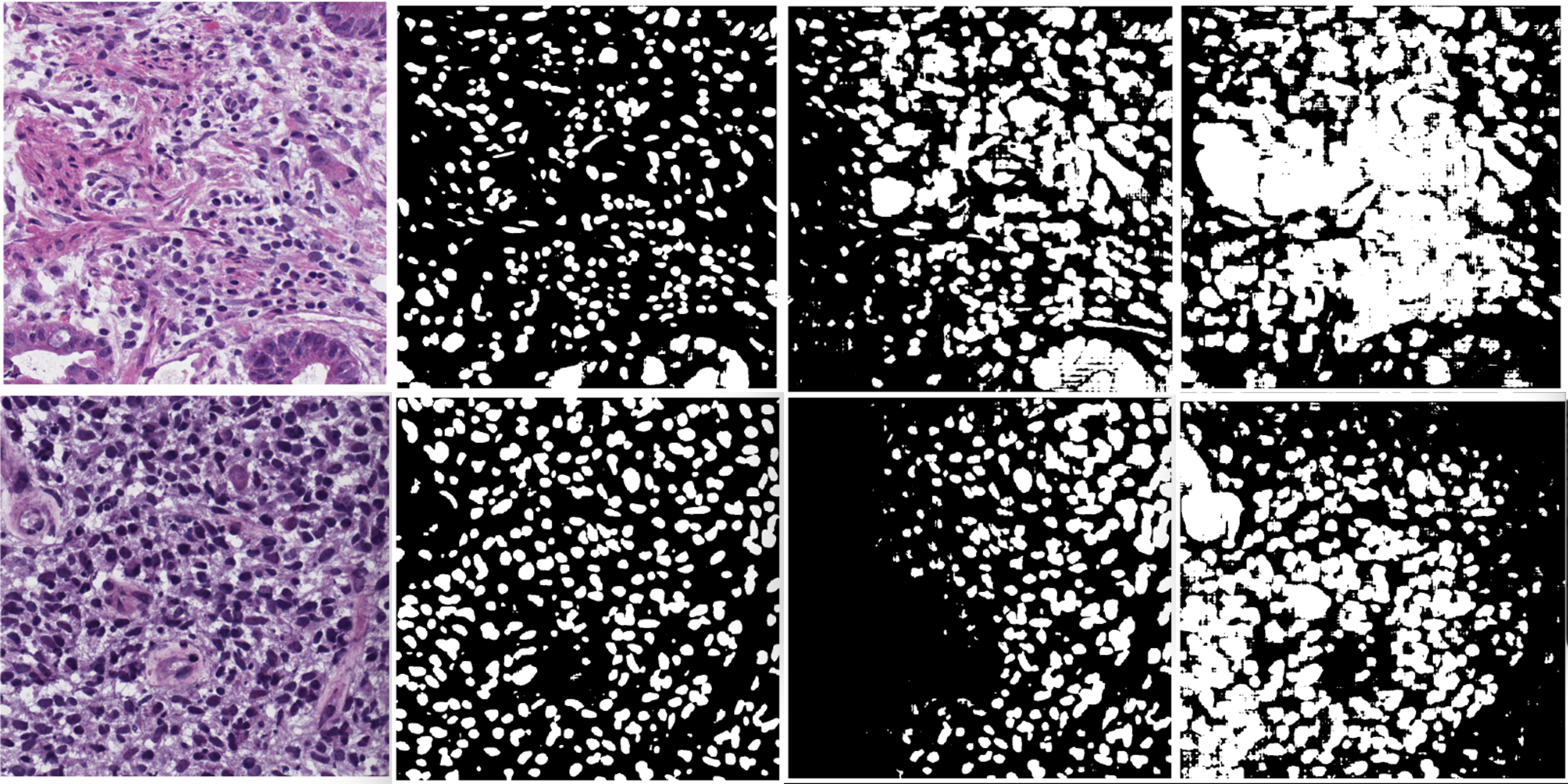}
    \caption{\label{fig:dexp} Two visualization examples of the segmentation results of SAM-FT from MoNuSeg dataset. The sequence of images displayed from left to right corresponds to the original histopathological specimen, followed by processed outputs with varying prompts of expert intervention: 0-expert, 1-expert, and 3-expert annotations, respectively.}
\end{figure}

We further evaluate the effects of our auto-prompts. In Table \ref{tab:decoder-results}, we utilize fine-tuned SAM without any auto-prompts, and then vary the number of expert prompts to check the performance on MoNuSeg and DSB. As observed, performance decreases when the number of manual expert prompts increases. This could be due to the model reverting to the original SAM behavior when fine-tuned without medical-specific prompts.
We also visualize two examples of the segmentation results in Figure \ref{fig:dexp} on MoNuSeg. As illustrated, increasing the number of expert prompts can inadvertently lead to the inclusion of extraneous areas, potentially compromising segmentation accuracy.

\subsubsection{Centroid-based prompt selection vs. Direct probability-based prompt selection}\label{sec:cvd}

As aforementioned, we assess the performance of different auto-prompt selection methods, which are centroid-based prompt selection and direct probability-based prompt selection. We evaluate the impact of selecting different numbers of points.
As shown in Table~\ref{tab:prompt_selection_comparison}, both $\mathcal{O}$ and $\mathcal{R}$ methods perform similarly when generating a small number of points (1-point and 3-point). However, as the number of points increases to 256, the centroid-based selection 
$\mathcal{O}$ significantly outperforms the direct probability-based selection $\mathcal{R}$ in terms of Dice and IoU metrics. This improvement comes at the cost of increased computation time, which is reported in the rightmost column in Table~\ref{tab:prompt_selection_comparison}. The larger number of points seems to benefit more from the spatial coherence that the centroid-based approach provides, leading to better segmentation performance despite the longer processing time.

\begin{table}[htbp]
\centering
\begin{tabular}{@{}lcccc@{}}
\toprule
\textbf{Method} & \textbf{F1} & \textbf{IoU} & \textbf{Dice} & \textbf{Time} $\nicefrac{s}{epoch}$ \\
\midrule
\(\mathcal{O}\)-1-point & 81.46 & 68.68 & 81.35 & 10 \\
\(\mathcal{O}\)-3-point & 81.57 & 68.79 & 81.46 & 10 \\
\(\mathcal{O}\)-256-point & \textbf{83.95} & \textbf{72.44} & \textbf{83.95} & \textbf{23} \\
\(\mathcal{R}\)-1-point & 81.57 & 68.76 & 81.4 & 10 \\
\(\mathcal{R}\)-3-point & 81.39 & 68.5 & 81.24 & 10 \\
\(\mathcal{R}\)-256-point & 81.73 & 69.04 & 81.6 & 12 \\
\bottomrule
\end{tabular}
\caption{\label{tab:prompt_selection_comparison}Comparison of different prompt selection methods with different point numbers. $\mathcal{O}$ refers to the centroid-based prompt selection method and $\mathcal{R}$ refers to the direct probability-based prompt selection method. ``-point'' represents different selected point number.}
\end{table}

\subsubsection{Efficiency Analysis}

\begin{figure}[htbp]
    \begin{center}
        \includegraphics[width=\linewidth]{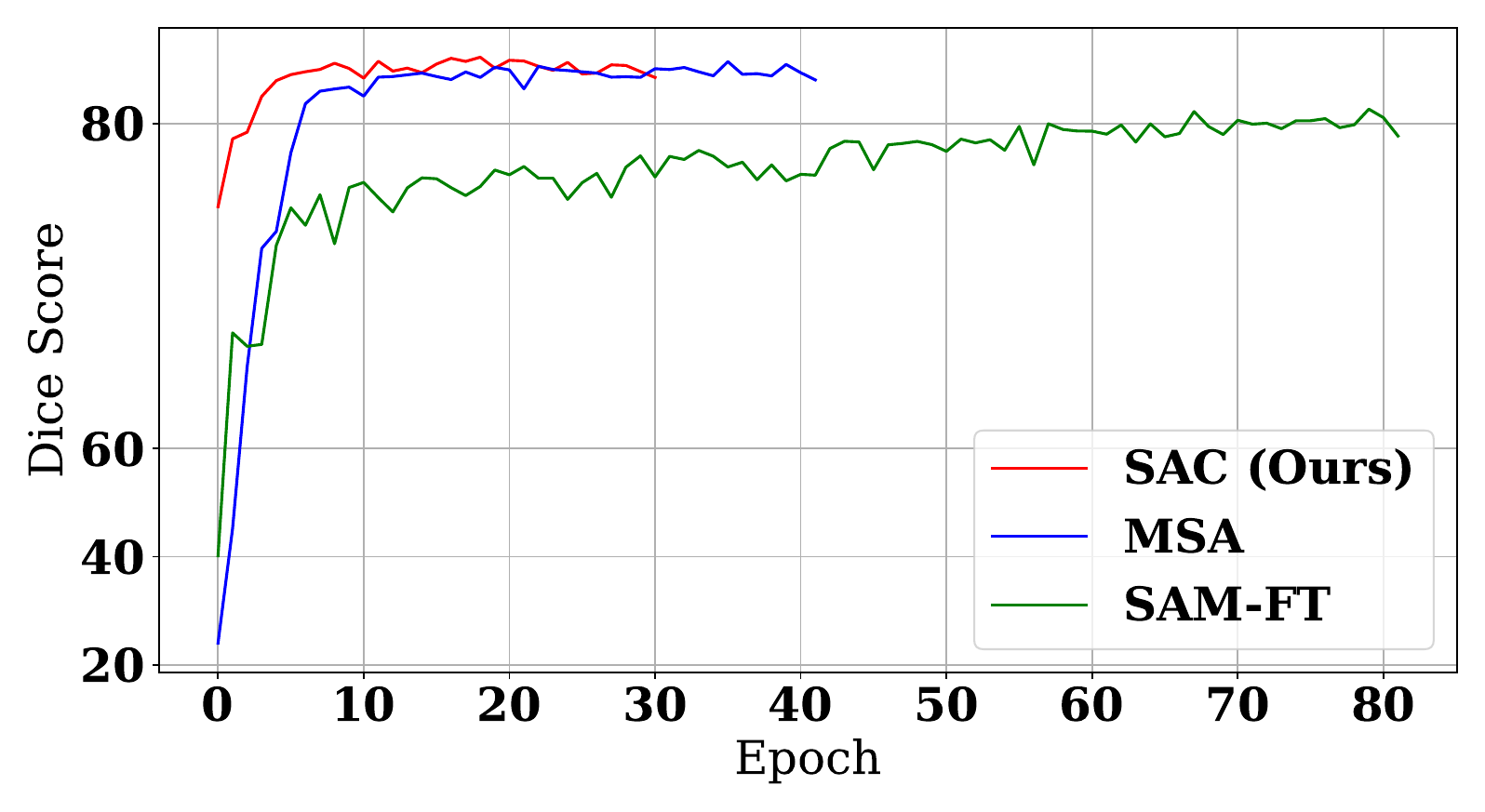}
    \end{center}
    \caption{Dice score convergence over epochs for SAM, MSA, and SAC (ours) on the MoNuSeg dataset.}
    \label{fig:dc}
\end{figure}

We conduct in-depth analysis to demonstrate that our SAC model is not only effective but also efficient. Figure \ref{fig:dc} shows the convergence of Dice scores across epochs for three different configurations: SAM-FT, MSA, and SAC (ours). Impressively, our SAC model exhibits a superior convergence, achieving the highest Dice scores in the fewest epochs. This indicates that the low-rank adaptations within our SAC framework significantly improve the model's learning efficiency. Conversely, the SAM-FT model, while initially improving, plateaus earlier, indicating potential overfitting or a lack of sufficient model complexity.

\begin{table}[htbp]
\centering
\begin{tabular}{@{}lccc@{}}
\toprule
\textbf{Model} & \textbf{Trainable} & \textbf{Freeze} & \textbf{Total} \\
\midrule
SAM-FT & 12,746,129 & 636,107,520 & 648,853,649 \\
MSA & 56,595,760 & 637,026,048 & 693,621,808 \\
SAC & 56,493,360 & 637,026,048 & 693,519,408 \\
\bottomrule
\end{tabular}
\caption{\label{tab:model_parameters}Comparison of parameter numbers.}
\end{table}

Table \ref{tab:model_parameters} compares the number of parameters across different models: SAM-FT, MSA, and SAC. SAM-FT serves as a base reference, which contains the least parameters since it is our backbone model. 
Compared with MSA, our model has fewer parameters, yet consistently achieves better segmentation performance. Furthermore, despite having more parameters, MSA converges more slowly and yields inferior results. 
Such observations suggest that the architecture and training regimen of SAC are more effective, indicating the potential for a better-tuned model for this task.

Due to space limitations, more ablation studies and additional experimental results are presented in the Supplementary Material \ref{suppl_Experiment_Detail}. Furthermore, a demo of our SAC pipeline can be found at \url{http://segment-any-cell.com}.






\section{Conclusion}
\label{sec:conclusion}

We introduce Segment Any Cell (SAC), an innovative auto-prompting fine-tuning framework based on SAM, specifically designed for nuclei segmentation tasks. Unlike other methods primarily focused on developing fine-tuning modules for foundation models like SAM, our approach highlights a crucial limitation: the need for accurate and informative prompts to fully exploit the capabilities of these models. To this end, we first develop a low-rank attention adapter within SAM to facilitate more efficient fine-tuning. Subsequently, we design an automatic prompt generator that creates precise prompts from medical images to guide the segmentation. Additionally, our framework accommodates a few-shot setting, allowing users to manually provide prompts for convenient use. Extensive experiment results validate the efficacy and superiority of SAC.
Moreover, our proposed framework can be easily adapted to other similar semantic segmentation tasks by altering the types of prompts, indicating its generalizability and flexibility.  
\newpage
\clearpage
\setcounter{page}{1}
\renewcommand{\thetitle}{}
\maketitlesupplementary

\setcounter{section}{0}
\renewcommand{\thesection}{\Alph{section}}
\renewcommand{\thesubsection}{\Alph{section}.\arabic{subsection}}

\section{Experiment Detail}
\label{suppl_Experiment_Detail}

\subsection{Baseline Models}
\label{suppl_baseline}
We have compared our proposed SAC with 12 baseline models, which are briefly described as follows:

\begin{itemize}
  \item U-Net~\cite{ronneberger2015u} is a well-established architecture that excels in preserving spatial information through its contracting and expansive paths.
  \item UCTransNet~\cite{wang2022uctransnet} addresses U-Net limitations by introducing Channel Transformers, including Channel-wise Cross Fusion Transformer (CCT) and Channel-wise Cross Attention (CCA) modules, significantly improving feature fusion and addressing semantic gap issues.
  \item MedT~\cite{valanarasu2021medical} is a transformer-based architecture tailored for medical image segmentation. 
  It tackles the limitations of convolutional neural networks by introducing a gated axial-attention mechanism with learnable gates for adaptive information control. The proposed Local-Global (LoGo) training strategy further enhances performance by operating on both global and local features.
  \item HistoSeg~\cite{wazir2022histoseg} combines Quick Attention Units in the encoder and decoder branches to enhance global and local feature representation. The method utilizes a multi-loss function, incorporating fixed focal loss, binary cross-entropy, and dice loss, focusing on challenging examples and precise boundary detection.
  \item DuAT~\cite{tang2022duat} (Dual-Aggregation Transformer Network) innovates medical image segmentation by introducing Global-to-Local Spatial Aggregation (GLSA) and Selective Boundary Aggregation (SBA) modules. GLSA captures global and local spatial features, aiding in identifying large and small objects. SBA enhances boundary details by selectively fusing low-level boundaries and high-level semantic information.
  \item SSFormer~\cite{wang2022stepwise} employs a pyramid Transformer encoder to enhance the model's generalization ability. It incorporates a Progressive Locality Decoder (PLD) as the decoder, designed for smoothing and effectively emphasizing local features within the Transformer, thereby improving the network's capability to process detailed information.
  \item MSRF-Net~\cite{srivastava2021msrf} introduces a Dual-Scale Dense Fusion (DSDF) block for medical image segmentation. This architecture excels in preserving high- and low-level features, improving segmentation performance.
  \item FANet~\cite{tomar2022fanet} uses a feedback attention mechanism, merging mask outputs from previous epochs with the current training epoch's feature map. This approach provides attention to learned features. The model supports iterative refinement of prediction masks during both training and testing, allowing it to adapt to sample variability. 
  \item MDM~\cite{pan2023masked} is a self-supervised representation learner derived from denoising diffusion models. In contrast to traditional diffusion models that utilize Gaussian noise, MDM employs a masking mechanism for scalability.
  \item DoubleU-Net~\cite{jha2020doubleu} innovatively enhances segmentation performance in medical images by incorporating two U-Net structures, a pre-trained VGG-19 model on ImageNet, and Atrous Spatial Pyramid Pooling (ASPP) to effectively handle diverse scale information.
  \item SAM~\cite{kirillov2023segment} is a foundation model for image segmentation, utilizing promptable tasks to generate real-time segmentation masks. Its architecture, consisting of image and prompt encoders along with a mask decoder, enables flexibility and interactivity. Trained on their introduced SA-1B dataset with over 1 billion high-quality masks, SAM demonstrates strong generalization and versatility in handling various segmentation tasks.
  \item MSA~\cite{wu2023medical} enhances SAM for medical image segmentation. Utilizing the Adaption technique, MSA integrates tailored Adapter modules to address SAM's limitations in the medical domain, considering factors like high dimensionality (3D) and unique visual prompts. It is important to note that MSA's results in cell segmentation are not presented. We compare the method frameworks of MSA and our approach.
\end{itemize}

\subsection{Loss Functions}\label{sec:lossc}

\subsubsection{Focal Loss}
The Focal Loss~\cite{lin2017focal} is designed to address class imbalance by focusing more on hard, misclassified examples:
\begin{equation}
    \text{Focal Loss} = -\alpha_t (1 - p_t)^\gamma \log(p_t),
\end{equation}
where \( p_t \) is the model's estimated probability for the class with label \( y=1 \), \( \alpha_t \) is a weighting term for balancing positive/negative examples, and \( \gamma \) is the focusing parameter.

\subsubsection{Dice Loss}
Dice Loss~\cite{milletari2016v} is commonly used in segmentation tasks to handle class imbalance, measuring the overlap between the prediction and the ground truth:
\begin{equation}
    \text{Dice Loss} = 1 - \frac{2 \sum_{i}^{N} p_i g_i}{\sum_{i}^{N} p_i^2 + \sum_{i}^{N} g_i^2},
\end{equation}
where \( p_i \) and \( g_i \) represent the predicted and ground truth values, respectively, for each element \( i \) in the N-dimensional space.

\subsubsection{DiceCELoss}
DiceCELoss combines Dice Loss with Cross-Entropy Loss, effective in segmentation tasks with class imbalance:
\begin{equation}
    \text{DiceCELoss} = 1 - \frac{2 \sum_{i}^{N} p_i g_i}{\sum_{i}^{N} p_i^2 + \sum_{i}^{N} g_i^2} - \sum_{i}^{N} g_i \log(p_i).
\end{equation}
This combines the overlap measure from Dice Loss with the pixel-wise classification accuracy from Cross-Entropy Loss.

\begin{figure*}[htbp]
    \centering
    \includegraphics[width=1\linewidth]{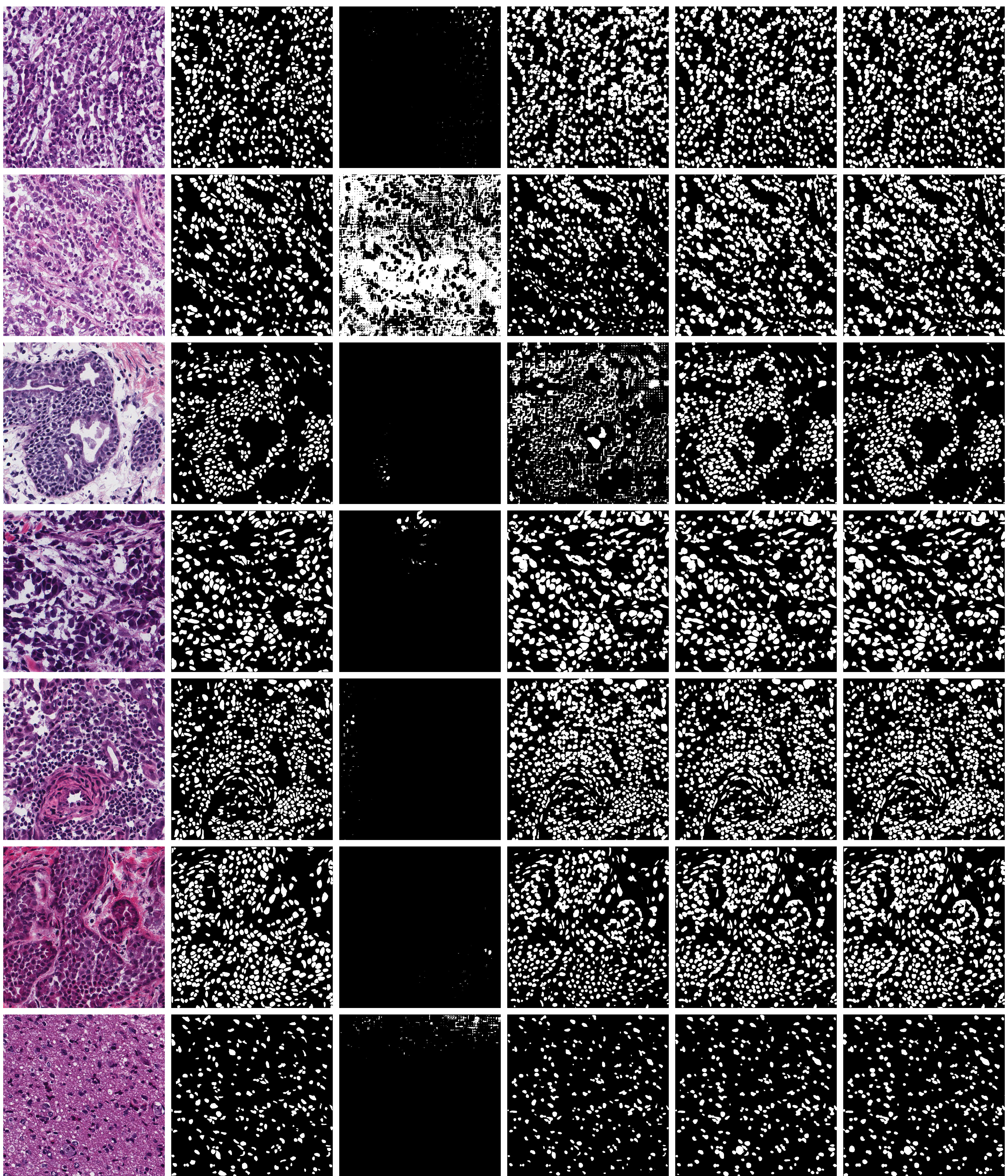}
    \vspace{2em}
\end{figure*}

\begin{figure*}[htbp]
    \centering
    \includegraphics[width=1\linewidth]{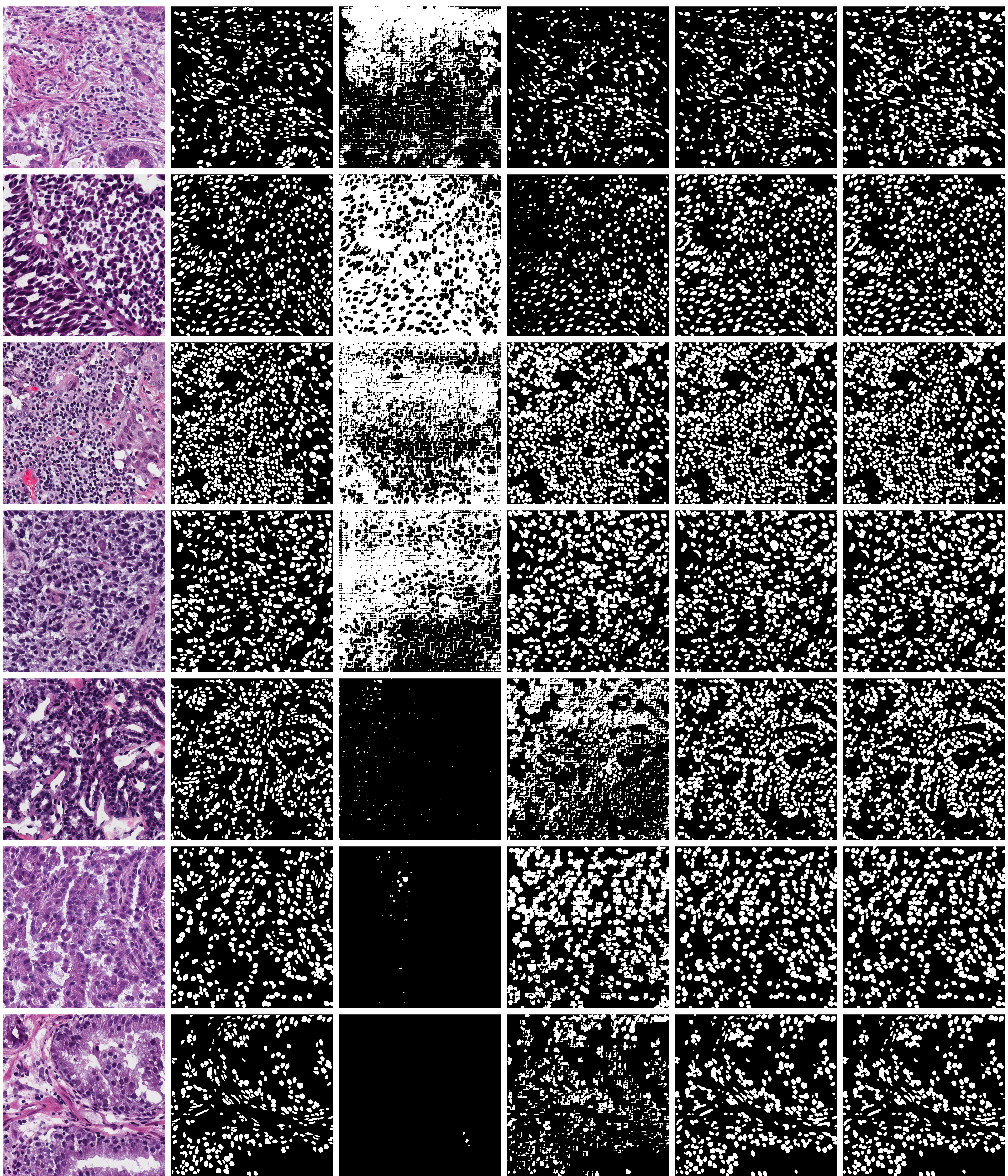}
    \caption{\label{fig:absa1} Illustration of 14 test images from MoNuSeg dataset. These images are segmented using only the auxiliary neural network with varying numbers of SAM prompts to demonstrate the effects of prompt quantity. From left to right: original image; segmentation mask; segmentation with 1 positive and 1 negative point; 3 positive and 3 negative points; 8 positive and 8 negative points; and 16 positive and 16 negative points.}
\end{figure*}

\subsection{Ablation studies}
\label{suppl_ab_study}

\subsubsection{Enhancement Through Incremental Prompt Amplification}

Figure \ref{fig:absa1} shows a transition from a basic segmentation to a more refined one as the number of points increases. This suggests that the auxiliary neural network is capable of refining its segmentation with an increasing number of prompts, which could indicate a more sophisticated understanding of the underlying features as more information is provided. Initially, with only one pair of prompts, the network may only grasp basic distinctions, but as more pairs are provided, the contours become more accurate, isolating regions of interest more effectively. The progression demonstrates the potential scalability and enhanced performance of the segmentation task with the incorporation of more prompts. Note that the prompt point pairs here are all generated by our auxiliary neural network without needing expert annotation.  For the complete demo, please visit \url{https://a31.segment-any-cell.com}.

\subsubsection{SAM VIT Backbone Comparison}

\begin{table}[h]
\centering
\begin{tabular}{lcccc}
\hline
\textbf{Backbone} & \textbf{F1 (\%)} & \textbf{IoU (\%)} & \textbf{Dice (\%)} & \textbf{Time (\nicefrac{s}{epoch})} \\
\hline
ViT-B & 83.56 & 71.79 & 83.52 & 4 \\
ViT-L & 83.90 & 72.35 & 83.89 & 8 \\
\textbf{ViT-H} & \textbf{84.03} & \textbf{72.61} & \textbf{84.11} & \textbf{10} \\
\hline
\end{tabular}
\caption{\label{tab:backbone_comparison}\textbf{Comparative results of SAM utilizing different backbones.}}
\end{table}

For the analysis of Table \ref{tab:backbone_comparison}, one can observe that as the complexity of the backbone increases from ViT-Base to ViT-Lage, and then to ViT-Huge \citep{dosovitskiy2020image}, there is a consistent improvement in the F1 score, IoU, and Dice coefficient. This suggests that a more complex backbone can capture finer details and achieve better segmentation results. However, this increased accuracy comes at the cost of computational efficiency, as indicated by the increase in time per epoch. The ViT-Huge backbone, while achieving the best segmentation performance, also requires the most extended time per epoch, which is more than double that of the ViT-Base backbone. This trade-off between accuracy and computational time is critical in determining the practicality of deploying such a model in real-world applications, where resources and time may be constrained.

\subsection{Additional Experiments}
\label{suppl_add_exps}

We have also applied our method to the non-cell segmentation dataset Gland Segmentation (GlaS)~\cite{sirinukunwattana2017gland}. The GlaS dataset focuses on gland segmentation and comprises 165 Hematoxylin and Eosin (H\&E) stained images, with 85 for training (37 benign, 48 malignant) and 80 for testing (37 benign, 43 malignant). The results are presented in Table \ref{tab:sac_glas_comparison}.

\begin{table}[h]
\centering
\begin{tabular}{lccc}
\hline
\textbf{Method} & \textbf{F1 (\%)} & \textbf{IoU (\%)} & \textbf{Dice (\%)} \\
\hline
U-Net~\cite{ronneberger2015u} & 77.78 & 65.34 & - \\
MedT~\cite{valanarasu2021medical} & 81.02 & 69.61 & - \\
UCTransNet~\cite{wang2022uctransnet} & - & 82.96 & 90.18 \\
MDM~\cite{pan2023masked} & - & 85.13 & 91.95 \\
MedT~\cite{valanarasu2021medical} & 81.02 & 69.61 & - \\
HistoSeg~\cite{wazir2022histoseg} & \textbf{98.07} & 76.73 & - \\
DoubleUnet~\cite{jha2020doubleu} & - & 80.3 & 89.07 \\
\textbf{SAC 0-expert} & 94.03 & \textbf{88.07} & \textbf{93.36} \\
\hline
\end{tabular}
\caption{\label{tab:sac_glas_comparison}Performance comparison of SAC with other methods on the GlaS dataset.}
\end{table}

In Table \ref{tab:sac_glas_comparison}, it is evident that the SAC 0-expert method demonstrates superior performance in terms of Dice score and IoU when compared to other methods, signifying a highly accurate segmentation outcome. 
The F1 coefficient is also impressive, although it is surpassed by the HistoSeg method.  
It is important to note that the HistoSeg method achieves an exceptionally high F1 score, yet its IoU is significantly lower than that of the SAC 0-expert, which may imply some overfitting to the F1 metric or an imbalance between precision and recall. 
The absence of data for certain metrics for some methods could imply that those particular metrics are not the focus or did not yield competitive results for those methods. 
The SAC 0-expert's balanced high performance across all three metrics underscores the robustness of our approach.

\newpage

{
    \small
    \bibliographystyle{ieeenat_fullname}
    \bibliography{main}
}


\end{document}